\documentclass[prd,preprint,superscriptaddress,showpacs]{revtex4}
\usepackage{graphicx}
\usepackage{latexsym}
\usepackage{amsmath}

\begin{document}

\title{Interactions in Dark Energy Models}
\author{Yi Zhang}
 \email{zhangyia@cqupt.edu.cn}
 \affiliation{Department of Astronomy, Beijing Normal university,  Beijing 100875, China}
\affiliation{College of Mathematics and Physics, Chongqing University of Posts and Telecommunications,
Chongqing 400065, China}
\author{Hui Li}
\email{lihui@ytu.edu.cn}
\affiliation{Department of Physics, Yantai University, Yantai 264005,
China}
\author{Yungui Gong}
\email{gongyg@cqupt.edu.cn}
 \affiliation{College of Mathematics and Physics, Chongqing University of Posts and Telecommunications,
Chongqing 400065, China}
\author{Zong-Hong
Zhu}
\email{zhuzh@bnu.edu.cn}
\affiliation{Department of Astronomy, Beijing Normal university,  Beijing 100875, China}

\begin{abstract}
We perform a full dynamical analysis by considering the interactions between dark energy and radiation, and dark energy and dark matter. We find that the interaction helps alleviate the coincidence problem for the quintessence model.
\end{abstract}

\preprint{arXiv: 1103.0718}

\maketitle
\section{Introduction}\label{sec1}

Recent observational data suggest a nearly flat universe with about $0.1\%$ relativistic matter,
30\% non-relativistic matter and 70\%  unknown energy which is called  dark energy \cite{Riess:1998cb}.
The most popular candidate for dynamical dark energy model
is the scalar field $\phi$ \cite{Liddle:1998jc}.
While these models are consistent with current observations,
the coincidence problem, which concerns the question why
the energy densities of dark energy and matter have the same order of magnitude
today even though they evolve very differently, is still the most important puzzle in cosmology.
Scalar field with  scaling behavior  could be used to alleviate the coincidence problem.
The scaling solution means that dark energy density scales with the scale factor,
so dark energy may track dark matter independent of the initial conditions.
However, the attractor of those models is totally dark energy domination $\Omega_\phi=1$.
By considering the interaction between dark energy and matter, it
is possible to get the accelerating attractor with constant ratio between the energy densities of dark
energy and dark matter, so the coincidence problem could be alleviated.

Although the radiation is negligible today,  it  played an important role in early universe.
The existence of radiation leads to the so-called  ``cosmic triple coincidence":
why the energy density of radiation today is three
orders of magnitude smaller than that of dark matter or dark energy
\cite{ArkaniHamed:2000tc}.  Borrowing the method of solving the first coincidence problem,
we consider the interaction between radiation and dark energy  to alleviate the ``cosmic triple coincidence" problem.
Although the Big Bang Nucleosynthesis (BBN) imposes a very strong constraint on radiation,
it is possible that a new kind of radiation called dark radiation may exist after BBN \cite{darkradiation}. Therefore,  it is possible to add an interaction between radiation  and dark energy after BBN.

To address the coincidence problem, we need to find accelerating attractor solution with order one ratio for $\Omega_m/\Omega_\phi$ because attractor solution is independent of initial conditions,
so we apply the dynamical analysis of phase-space with the introduction of interactions between dark energy, dark matter and radiation.
The phase-space analysis is a standard method to study the stability of the fixed points \cite{dynamics,threefluid,quintessenced,phantomd,Zhang:2010icb}.

The letter is organized as follows. In section \ref{sec2}, we introduce our dark energy model.
In section  \ref{sec3},  the results of
phase-space analysis are obtained.
Conclusion is presented in section  \ref{sec4}.

\section{The interactions between cosmic components}\label{sec2}

Based on the  flat Friedmann-Robertson-Walker  metric,
\begin{eqnarray}
 ds^{2}=-dt^{2}+a^{2}(t)\sum^{3}_{i=1}(dx^{i})^{2},
\end{eqnarray}
where $a$ is the scale factor, Friedmann equation is
  \begin{equation}
 H^{2}=\frac{1}{3m_{pl}^{2}}(\rho_{\phi}+\rho_{r}+\rho_{m}),
 \end{equation}
where $H=\dot{a}/a$ is Hubble parameter, $m_{pl}^2=(8\pi G)^{-1}$, $\rho_\phi$, $\rho_m$,
and $\rho_r$ denote the energy densities of
dark energy which is assumed to be a scalar field,  pressureless matter, and  radiation, respectively.
Writing $\rho_{tot}=\rho_\phi+\rho_m+\rho_r$, the equation of energy conservation is,
 \begin{equation}
 \label{etotconsv}
 \dot{\rho}_{tot}+3H(\rho_{tot}+p_{tot})
=0.
\end{equation}

If there is no interaction between dark energy and the other field in the universe,
then it is almost impossible to detect dark energy through standard technique without gravitational effect involved.
Therefore, we consider the interactions between dark energy and the other matter components in the universe.

For the interaction between dark energy and radiation, we assume that,
 \begin{equation}
 \label{rc}
 \dot{\rho_{r}}+4H\rho_{r}=\Gamma \dot{\phi}^{2},
 \end{equation}
 where the interaction term $\Gamma
\dot{\phi}^{2}$ is the source of  radiation and $\Gamma $ is the dissipative coefficient which is related to
the microscopic physics of the interaction. This interaction form was first
proposed in  warm inflation \cite{Berera:1995wh,Berera:1995ie,warm,Zhang:2009ge,Hall:2003zp} based on supersymmetry (SUSY).
If $\Gamma>0$, the energy transfer from dark energy to radiation makes the radiation
to decease slower and dark energy to decrease faster, therefore provides the possibility to alleviate the coincidence problem.

Motivated by the different forms of the dissipative
coefficients in Refs.
\cite{Berera:1995wh,Berera:1995ie,warm,Zhang:2009ge,Hall:2003zp},
a general form of $\Gamma$ was proposed in Ref. \cite{Zhang:2009ge},
 \begin{equation}
\label{gf}
\Gamma=C_{\phi}\frac{T^{\tilde{m}}}{\phi^{\tilde{m}-1}},
 \end{equation}
where $C_{\phi}$ and $\tilde{m}$ are parameters related to the dissipative microscopic
dynamics. Moreover,
a more general phenomenological form of $\Gamma$ was proposed  in Ref. \cite{Hall:2003zp},
 \begin{equation}
 \label{general}
 \Gamma=C_{\phi}\frac{T^{m}}{\phi^{n}},
 \end{equation}
 where $m$ and $n$ are constants.
To be more realistic, the interaction between dark energy and radiation is assumed
to be turned on after the BBN period.

For the interaction between dark energy and matter, we consider the interaction form which is proportional to $\rho_m\dot\phi$,
the energy conservation equation of $\rho_m$ is
 \begin{equation}
 \label{emconsv}
 \dot{\rho}_{m}+3H\rho_{m}=\frac{\alpha\rho_{m}\dot{\phi}}{m_{pl}},
 \end{equation}
the interaction form on the right hand side was motivated from Brans-Dicke theory
and  is widely used to alleviate the coincidence problem \cite{Damour:1994ya}.
If $\alpha>0$, the energy transfer from dark energy to dark matter makes dark matter
to decease slower and dark energy to decrease faster. Therefore, the interactions based on Eqs. (\ref{rc}) and (\ref{emconsv}) provide the possibility of alleviating the ``cosmic triple coincidence" problem.

There are evidences that the equation of state (EoS) parameter of dark energy crosses over
the value of $-1$ \cite{Riess:2004nr}, so we consider both quintessence and phantom cases.
The energy density and pressure of dark energy are written as
\begin{eqnarray}
&&\rho_{\phi}=\epsilon \frac{\dot{\phi}^{2}}{2}+V(\phi),\\
&&p_{\phi}=\epsilon\frac{\dot{\phi}^{2}}{2}-V(\phi),
\end{eqnarray}
where the sign $\epsilon=\pm 1$. If $\epsilon=1$, then the scalar field is
the quintessence field; if $\epsilon=-1$, then the scalar field is the  phantom field.

Combining  the energy conservation Eqs. (\ref{etotconsv}), (\ref{rc}) and (\ref{emconsv}), the equation of motion
of the scalar field becomes
\begin{equation}
\label{me}
\ddot{\phi}+(3H+\Gamma)\dot{\phi}+\epsilon V_{,\phi}=-\epsilon\frac{\alpha\rho_{m}}{m_{pl}},
\end{equation}
where $V_{,\phi}$ means the partial derivative of the potential $V$ with respect to the scalar field $\phi$.

The exponential potential is one of the most often used potential in
cosmology \cite{expmoti,wandsexponential,coincidence}.  In higher-dimensional gravitational theories such as
superstring and Kaluza-Klein theories, exponential
potentials are often obtained by the compactification of  the internal spaces of extra dimensions.
Moreover, it is known that exponential potentials can arise from
gaugino condensation as a non-perturbative effect, and from supergravity corrections in global supersymmetric
theories. The possible role of exponential potentials used to solve
the coincidence problem was studied extensively, and the scaling solution was found for the exponential potential.
In this letter, we consider the exponential potential for the scalar field,
 \begin{equation}
 \label{exp}
V=V_{0}e^{-\lambda\phi/m_{pl}},
  \end{equation}
where $\lambda$ and $V_{0}$ are the model parameters. If the scaling solution still exists in the interacting model,
then the ``triple coincidence problem" may be alleviated.
In the following, to investigate the coincidence problem, we present a phase-space analysis.

\section{The Phase-space Analysis}\label{sec3}

We define the dimensionless parameters as
\begin{eqnarray}
\nonumber&&x=\frac{\dot{\phi}}{\sqrt{6}m_{pl}H},\,\,\,\,y=\frac{\sqrt{V}}{\sqrt{3}m_{pl}H},\,\,\,\,z=\frac{\sqrt{6}m_{pl}}{\phi},\\
&&u=\frac{\sqrt{\rho_{r}}}{\sqrt{3}m_{pl}H},\,\,\,\,v=\frac{\sqrt{\rho_{m}}}{\sqrt{3}m_{pl}H}.
\end{eqnarray}
By using these dimensionless parameters, we get $\Omega_m=v^2$, $\Omega_r=u^2$, $\Omega_\phi=\epsilon x^2+y^2$
and $w_\phi=(\epsilon x^2-y^2)/(\epsilon x^2+y^2)$.
The dynamical evolution of the universe becomes
\begin{eqnarray}
\label{primex}
&&x'=-x\frac{\dot{H}}{H^{2}}-\frac{(3H+\Gamma)\dot{\phi}+\epsilon V_{,\phi}+\epsilon\alpha\rho_{m}/m_{pl}}{\sqrt{6}m_{pl}H^{2}},\\
\label{primey}
&&y'=-y\frac{\dot{H}}{H^{2}}+\frac{\dot{\phi}}{\sqrt{6}m_{pl}H}\frac{V_{,\phi}}{\sqrt{2V}H},\\
\label{primez}
&&z'=-\frac{6m_{pl}^{2}}{\phi^{2}}\frac{\dot{\phi}}{\sqrt{6}m_{pl}H},\\
\label{primeu}
 &&u'=-u\frac{\dot{H}}{H^{2}}-2u+\frac{\Gamma\dot{\phi}^{2}}{2\sqrt{3\rho_{r}}m_{pl}H^{2}},\\
 \label{primev}
 &&v'=-v\frac{\dot{H}}{H^{2}}-\frac{3}{2}v+\frac{\alpha\sqrt{\rho_{m}}\dot{\phi}}{2\sqrt{3}(m_{pl}^{2}H^{2})}.
\end{eqnarray}
where a prime means the derivative with respect to $\ln a$.

In addition, the evolution of the Hubble parameter could be expressed as
 \begin{equation}
 -\frac{\dot{H}}{H^{2}}=3\epsilon x^{2}+2u^{2}+\frac{3}{2}v^{2}.
  \end{equation}
After substituting the above equation into the dynamical Eqs. (\ref{primex})-(\ref{primev}) of the dimensionless parameters,
the  following term becomes the only one that needs to be expressed as the dimensionless parameter,
 \begin{equation}
 \frac{\Gamma}{H}=\frac{C_{\phi}}{H}(\frac{\sqrt{3}m_{pl}H u}{\sqrt{C_{r}}})^{m/2}(\frac{z}{\sqrt{6}m_{pl}})^{n},
 \end{equation}
where $C_r=\pi^2 g^*/30$ and $g^*$ is the effective number of degrees of freedom for
the relativistic species in the universe. If $m=2$, the dynamical system could be written as an autonomous system,
and the evolutions of the dimensionless parameters are
\begin{eqnarray}
\label{x1}
&&x'=-3x+\epsilon\lambda\sqrt{\frac{3}{2}}y^{2}+x(\epsilon3x^{2}+2u^{2}+\frac{3}{2}v^{2})-K x z^{n} u-\epsilon\alpha\sqrt{\frac{3}{2}}v^{2}, \\
\label{y1}
&&y'=-\lambda\sqrt{\frac{3}{2}}x y+y(\epsilon3x^{2}+ 2u^{2}+\frac{3}{2}v^{2}), \\
\label{z1}
 &&z'=-z^{2}x,\\
\label{u1}
&&u'=u(\epsilon3x^{2}+2u^{2}+\frac{3}{2}v^{2})-2u+Kx^{2}z^{n},\\
\label{v1}
&&v'=v(\epsilon3x^{2}+2u^{2}+\frac{3}{2}v^{2})-\frac{3}{2}v+\alpha\sqrt{\frac{3}{2}}xv.
\end{eqnarray}
where $K=-\sqrt{2C_{r}}(6m_{pl}^{2})^{(n-1)/2}/C_{\phi}$.
The critical points of the autonomous system could be derived by
imposing the conditions $x'=y'=z'=u'=v'=0$.
Nevertheless, since Friedman equation is
 \begin{equation}
 \epsilon x^{2}+y^{2}+u^{2}+v^{2}=1,
 \end{equation}
 which acts as a constraint equation, there are only four independent variables,
 we choose them as $x$, $y$, $z$ and $u$.
 After obtaining the fixed points, we perturb the four independent dynamical equations of
Eqs. (\ref{x1})-(\ref{v1}) around the fixed points, and then linearize the perturbed equations,
the four eigenvalues of the coefficient matrix of the linearized
equations determine the stability of the corresponding critical
point.  When the real parts of the eigenvalues  are all positive, the corresponding critical
point is an unstable fixed point. When the real parts of the eigenvalues  are all negative, the
corresponding critical point is a stable attractor point. When some of the real parts of the
eigenvalues  are negative, and some of the real parts of the eigenvalues are positive, the
corresponding critical point is an unstable saddle point \cite{dynamics}.

\subsection{The Quintessence Case }
\begin{figure}
  \center
  \includegraphics[width=0.6\textwidth]{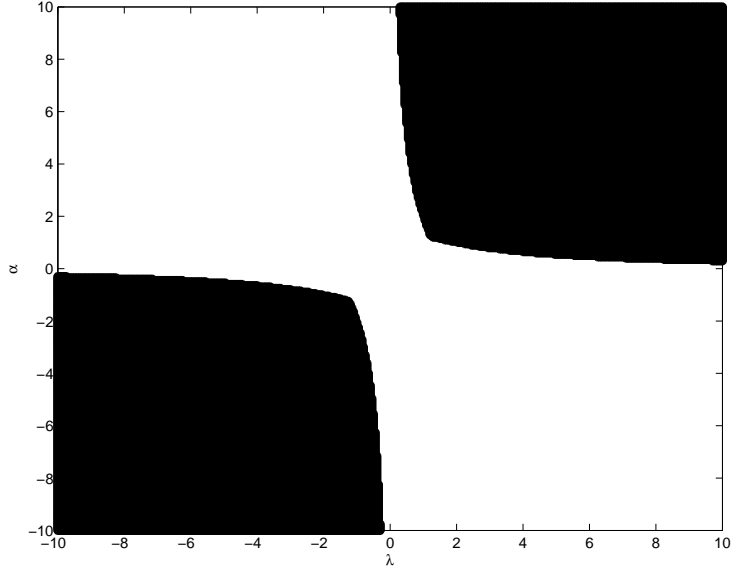}
    \caption{\label{sqm1} The parameter space for the point $S_{qM}$ to be stable. }
\end{figure}

\begin{figure}
\center
  \includegraphics[width=0.6\textwidth]{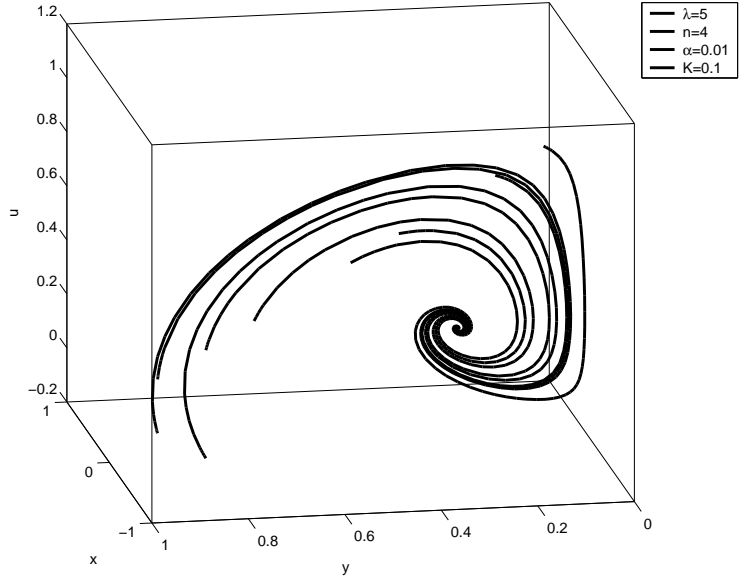}
    \caption{\label{SqM} The phase-space trajectories with $\lambda=5$, $\alpha=0.01$, $n=4$, $K=0.1$, the stable fixed point is $S_{qM}$.}
\end{figure}

 In the quintessence case, the EoS parameter  could be expressed as
 \begin{equation}
 \omega_{\phi}=\frac{\dot{\phi}^{2}/2-V}{\dot{\phi}^{2}/2+V}=\frac{x^{2}-y^{2}}{x^{2}+y^{2}}.
 \end{equation}
Based on the dynamical Eqs. (\ref{x1})-(\ref{v1}),  we derive the critical points and
list their properties in  TABLE \ref{tab}. When radiation is absent, we recover the results given
by Boehmer etal. \cite{quintessenced}.
The critical points are explained as follows.
\begin{description}
\item[$R_{q}$~~]The radiation dominated phase;\\
For this fixed point, $\Omega_r=1$, the eigenvalues are $\lambda_{1}=1$, $\lambda_{2}=2$, $\lambda_{3}=0$, and $\lambda_{4}=-1-Kz^{n}$. Because $\lambda_1>0$
and $\lambda_2>0$, it is an unstable fixed point. In particular,
if $\lambda_4<0$, then $R_q$ is an unstable saddle point.
Therefore, the universe does not always stay radiation dominated phase and it will go to matter dominated phase.

\item[$S_{qR}$~]The deceleration phase with radiation and scalar field;\\
For this fixed point, $\Omega_r=1-4/\lambda^2$, $\Omega_\phi=4/\lambda^2$ and $w_\phi=1/3$.
The scalar field tracks the dynamics of  radiation, with a constant ratio between the two energy densities. For
this critical point, we choose $x$, $y$, $z$, $v$ as the independent dynamical parameters. The eigenvalues are $\lambda_{1}=0$,  $\lambda_{2}=1/2+2\alpha/\lambda$, $\lambda_{3}=-1/2+\sqrt{-15/4+16/\lambda^{2}}$, and $\lambda_{4}=-1/2-\sqrt{-15/4+16/\lambda^{2}}$.
If $\alpha/\lambda>-1/4$ or $\lambda^{2}<4$, the fixed point is an unstable saddle point. Since
the fixed point is unstable and $\Omega_\phi/\Omega_r$ is a constant, when
$\alpha/\lambda>-1/4$ and $\lambda^2>4$, once the universe exits from this phase
with constant $\Omega_r/\Omega_\phi$ and enters into
the accelerating phase, the coincidence problem becomes weakened.

\item[$M_{q}$~~] The deceleration phase with matter and scalar field; \\
For this fixed point, $\Omega_m=1-2\alpha^2/3$, $\Omega_\phi=2\alpha^2/3$ and $w_\phi=1$.
When the interaction is absent, $\alpha=0$, it becomes the matter dominated phase and it
is an unstable saddle point. If the interaction exists, the ratio $\Omega_\phi/\Omega_m$ is
a constant, but there is no acceleration since $w_\phi=1$.
The eigenvalues are $\lambda_{1}=0$, $\lambda_{2}=\alpha^{2}-1/2$, $\lambda_{3}=-3/2+\alpha^{2}$, and $\lambda_{4}=\lambda\alpha+\alpha^{2}+3/2$.
If $\alpha^{2}>1/2$, or $\lambda^2<6$, or $\lambda^2\ge 6$ and $(-\lambda+\sqrt{\lambda^2-6})/2<\alpha<\sqrt{6}/2$, this point is an unstable fixed point. In this phase, the scalar field
cannot represent dark energy, but it can provide the
correct order of magnitude for the ratio $\Omega_\phi/\Omega_m$. So if the universe enters the dark energy
dominated phase after this phase, the coincidence problem becomes weakened, i.e., the interaction helps weaken
the coincidence problem.

\item[$S_{qRM}$] The deceleration phase with matter and radiation;\\
For this fixed point, $\Omega_m=1/3\alpha^{2}$, $\Omega_r=1-1/2\alpha^2$, $\Omega_{\phi}=1/6\alpha^{2}$
and $w_\phi=1$. Since $\alpha^2>1/2$, this phase requires strong coupling between dark sectors, $\Omega_m/\Omega_\phi=2$ and $\Omega_\phi/\Omega_r=1/(6\alpha^2-3)$. The eigenvalues are $\lambda_{1}=0$, $\lambda_{2}=2+\lambda/2\alpha$, $\lambda_{3}=-1/2+\sqrt{-3/4+1/2\alpha^{2}}$ and $\lambda_{4}=-1/2-\sqrt{-3/4+1/2\alpha^{2}}$. When $\lambda/\alpha>-4$, this point is an unstable
fixed point, so the universe may exit from this phase and enter into the accelerating phase.
Because of nonzero values of $\Omega_\phi/\Omega_m$ and $\Omega_\phi/\Omega_r$,
the interaction helps weaken the triple coincidence problem.

\item[$K_{q+}$~]The kinetic term dominated phase A;\\
For this fixed point, $\Omega_\phi=1$ and $w_\phi=1$, it is a deceleration phase which is
not physically interesting. The eigenvalues are  $\lambda_{1}=1$, $\lambda_{2}=0$, $\lambda_{3}=3-\lambda\sqrt{3/2}$, and $\lambda_{4}=3+2\alpha\sqrt{3/2}$,
so this point is  an unstable fixed point. In particular,
when $\alpha>-\sqrt{3/2}$ and $\lambda<\sqrt{6}$, all the eigenvalues are positive.

\item[$K_{q-}$~]The kinetic term dominated phase B;\\
For this fixed point, $\Omega_\phi=1$ and $w_\phi=1$, it is also a deceleration phase which is
not physically interesting.
The eigenvalues are $\lambda_{1}=1$, and $\lambda_{2}=0$, $\lambda_{3}=3+\lambda\sqrt{3/2}$, and $\lambda_{4}=3-2\alpha\sqrt{3/2}$, so this point is  an unstable fixed point as well. In particular,
when $\alpha<\sqrt{3/2}$ and $\lambda>-\sqrt{6}$, all the eigenvalues are positive.

\item[$S_{q}$~~] The dark energy dominated phase;\\
For this fixed point, $\Omega_\phi=1$ and $w_\phi=-1+\lambda^2/18$.
The eigenvalues are $\lambda_{1}=0$, $\lambda_{2}=-2+\lambda^{2}/2$, $\lambda_{3}=-3+\lambda^{2}/2$ and $\lambda_{4}=-3+\lambda^{2}+\lambda\alpha$. The stability condition requires that all the eigenvalues are negative which means $\lambda^{2}<4$ and $\lambda \alpha<(3-\lambda^{2})$. However, this attractor
does not solve the coincidence problem.

\item[$S_{qM}$~]The scaling solution with  nonzero $\Omega_m$ and $\Omega_\phi$;\\
For this fixed point, $\Omega_m=(\lambda(\lambda+\alpha)-3)/(\lambda+\alpha)^2$,
$w_\phi=-\alpha(\lambda+\alpha)/(\alpha(\lambda+\alpha)+3)$, and $\Omega_m/\Omega_\phi$ is a constant.
The existence condition of the fixed point is $\alpha(\lambda+\alpha)>-3/2$
and $\lambda(\lambda+\alpha)>3$ (hereafter Condition 1).
The eigenvalues are $\lambda_{1}=0$, $\lambda_{2}=-1/2-3\alpha/[2(\lambda+\alpha)]$, $\lambda_{3}=-3/2-3\alpha/2(\lambda+\alpha)+a$ and $\lambda_{4}=-3/2-3\alpha/2(\lambda+\alpha)-a$, where $a=\sqrt{9/16+9\alpha^{2}/16(\alpha+\lambda)^{2}+81\alpha/8(\lambda+\alpha)+27/2(\lambda+\alpha)^{2}-9\lambda/2(\lambda+\alpha)-3\lambda\alpha}$. If there is no interaction between the dark sectors,
we recover the standard tracking solution $w_\phi=w_m=0$ \cite{wandsexponential}. If $\alpha(\alpha+\lambda)\gg 3$, $w_\phi\sim -1$,
we get a dark energy dominated solution. In FIG. \ref{sqm1}, we plot the regions of the parameters $\alpha$
and $\lambda$ that make the fixed point to be stable. To illustrate the stability of the fixed point, we
plot the phase trajectories with $\lambda=5$, $\alpha=0.01$, $n=4$, $K=0.1$ in FIG. \ref{SqM}.
By choosing appropriate parameters, we get dark energy dominated phase with $\Omega_m/\Omega_\phi\approx 3/7$,
so the coincidence problem can be alleviated.

\end{description}

For the quintessence case,  we have several physical interesting unstable fixed points. Immediately after
the total radiation dominated phase $R_q$, the phase $S_{qR}$ provides the possibility of nonzero $\Omega_\phi$ at early time
and helps weaken the coincidence problem, then the phase $M_q$ provides the matter domination with small $\Omega_\phi$
which also helps weaken the coincidence problem, it follows by the phase $S_{qRM}$ which helps weaken the triple
coincidence problem, finally the universe enter the dark energy dominated phase $S_{qM}$ with the observed
ratio between dark energy and dark matter.

\begin{table*}[t]
\begin{center}
\begin{tabular}{|c|c|c|c|c|c|}
\hline
 &$(x,y,z,u,v)$ & existence& $w_{\phi}$& $\Omega_{\phi}$& Acceleration\\
\hline
\hline  $R_{q}$&$(0,0,z,1,0)$ &all $\lambda$& $0$& 0& No\\
\hline  $M_{q}$&$(-\sqrt{\frac{2}{3}}\alpha,0,0,0,\sqrt{1-\frac{2\alpha^{2}}{3}})$ & $\alpha^{2}<\frac{3}{2}$& $1$  & $2\alpha^{2}/3$& No \\
\hline  $K_{q+}$&$(1,0,0,0,0)$& all $\lambda$& $1$   & $1$& No \\
 \hline  $K_{q-}$&$(-1,0,0,0,0)$& all $\lambda$& $1$ & $1$ & No \\
\hline  $S_{q}$&$(\frac{\lambda}{\sqrt{6}},\sqrt{1-\frac{\lambda^{2}}{6}},
0,0,0)$&$\lambda^{2}<6$&$-1+\frac{\lambda^{2}}{18}$ & $1$ & $\lambda^{2}<6$ \\
\hline  $S_{qM}$&$(\sqrt{\frac{3}{2(\lambda+\alpha)^{2}}},\sqrt{\frac{2\alpha(\lambda+\alpha)+3}{2(\lambda+\alpha)^{2}}},0, 0, \sqrt{\frac{\lambda(\lambda+\alpha)-3}{(\lambda+\alpha)^{2}}})$& Condition 1 &$-\frac{\alpha(\lambda+\alpha)}{\alpha(\lambda+\alpha)+3}$ & $\frac{3+\alpha(\alpha+\lambda)}{(\lambda+\alpha)^{2}}$ & $\frac{1}{\alpha(\lambda+\alpha)}<\frac{2}{3}$ \\
\hline  $S_{qRM}$&$(-\frac{1}{\sqrt{6}\alpha},0,0,\sqrt{1-\frac{1}{2\alpha^{2}}},\frac{1}{\sqrt{3}\alpha})$& $\alpha^{2}>\frac{1}{2}$ & $1$& $\frac{1}{6\alpha^{2}}$ & No \\
\hline $S_{qR}$& $((\frac{8}{3\lambda^{2}})^{1/2},(\frac{4}{3\lambda^{2}})^{1/2},0, \sqrt{1-\frac{4}{\lambda^{2}}},0)$&$\lambda^{2}>4$&$1/3$ & $\frac{4}{\lambda^{2}}$ & No \\
\hline
\end{tabular}
\end{center}
\caption[crit]{\label{tab}
 The properties of the critical points when $m=2$ in the quintessence case.}
\end{table*}

\subsection{The Phantom Case}
For the phantom case, the EoS parameter of the scalar field  is
\begin{equation}
 \omega_{\phi}=\frac{-\dot{\phi}^{2}/2-V}{-\dot{\phi}^{2}/2+V}=\frac{-x^{2}-y^{2}}{-x^{2}+y^{2}}.
 \end{equation}
The properties of the critical points are summarized in  TABLE \ref{tab2}.
When radiation is absent, we recover the results given by Leon and Saridakis \cite{phantomd}.

\begin{description}
\item[$R_{p}$~~]The radiation dominated phase;\\
For this fixed point, $\Omega_r=1$ and $w_{\phi}=1$. The eigenvalues are $\lambda_{1}=1,\lambda_{2}=2,\lambda_{3}=0,\lambda_{4}=-1-Kz^{n} $, so this point
is  an unstable fixed point.

\item[$M_{p}$~~] The matter dominated phase;\\
For this fixed point, $\Omega_{\phi}=-2\alpha^{2}/3$ and $w_{\phi}=1$.
The eigenvalues are $\lambda_{1}=0$, $\lambda_{2}=-\alpha^{2}-1/2$, $\lambda_{3}=-3/2-2\alpha^{2}-3\alpha^{2}$, and $\lambda_{4}=\lambda\alpha-\alpha^{2}+3/2$. This point exists only when $\alpha=0$, and it is an unstable fixed point.

\item[$S_{p}$~~]The phantom dominated phase;\\
For this fixed point, $\Omega_{\phi}=1$ and $w_{\phi}=-\lambda^{2}/18-1$.
The eigenvalues are $\lambda_{1}=0$, $\lambda_{2}=-2-\lambda^{2}/2$, $\lambda_{3}=-3-\lambda^{2}-\alpha\lambda$ and $\lambda_{4}=-\lambda^{2}/2-3$. If $\alpha^2<12$, or $\alpha^2>12$ and $\lambda>(-\alpha+\sqrt{\alpha^2-12})/2$
or $\lambda<(-\alpha-\sqrt{\alpha^2-12})/2$, $\lambda_3<0$, the point is a stable fixed point.
Since $\Omega_\phi=1$ does not solve the coincidence problem, we do not discuss the point in detail.

\item[$S_{pM}$~~]The scaling phase with acceleration;\\
For this fixed point, $\Omega_{\phi}=(-3+\alpha(\lambda+\alpha))/(\lambda+\alpha^{2})$,
$w_{\phi}=-\alpha(\lambda+\alpha)/(-3+\alpha(\lambda+\alpha))$ and $\Omega_m=1-\Omega_\phi$.
The existence condition of the fixed point is $\alpha(\lambda+\alpha)>3$
and $\lambda(\lambda+\alpha)>-3$ (hereafter Condition 2).
The eigenvalues are $\lambda_{1}=0$, $\lambda_{2}=-1/2-3\alpha/2(\lambda+\alpha)$, $\lambda_{3}=-3/2-3\alpha/2(\lambda+\alpha)+b $ and $\lambda_{4}=-3/2-3\alpha/2(\lambda+\alpha)-b$, where $b=\sqrt{9/16+3\alpha\lambda-27/2(\lambda+\alpha)^{2}-9\lambda/2(\lambda+\alpha)+9\alpha^{2}/16(\alpha+\lambda)^{2}+81\alpha/8(\alpha+\lambda)}$. The point is an unstable fixed point.
\end{description}

Therefore, the interacting phantom model cannot alleviate the coincidence problem.

\begin{table*}[t]
\begin{center}
\begin{tabular}{|c|c|c|c|c|c|}
\hline
 &$(x,y,z,u,v)$ & existence& $w_{\phi}$ &$\Omega_{\phi}$& Acceleration\\
\hline
\hline  $R_{p}$&$(0,0,z,1,0)$ &All $\lambda$& $1$ &$0$ & No \\
\hline  $M_{p}$&$(\sqrt{\frac{2}{3}}\alpha,0,0,0,\sqrt{1+\frac{2\alpha^{2}}{3}})$ & $\alpha=0$& $1$ & $-\frac{2}{3}\alpha^{2}$ & No\\
\hline  $S_{p}$&$(-\frac{\lambda}{\sqrt{6}},\sqrt{1+\frac{\lambda^{2}}{6}},
0,0,0)$&All $\lambda$& $-\frac{\lambda^{2}}{18}-1$ & $1$ & All $\lambda$\\
\hline  $S_{pM}$&$(\frac{\sqrt{6}}{2(\lambda+\alpha)}, \sqrt{\frac{-3+2\alpha(\lambda+\alpha)}{2(\lambda+\alpha)^{2}}},0,0,\sqrt{1+\frac{3-\alpha(\lambda+\alpha)}{(\lambda+\alpha)^{2}}})$
& Condition 2& $\frac{-\alpha(\lambda+\alpha)}{-3+\alpha(\lambda+\alpha)}$& $\frac{-3+\alpha(\lambda+\alpha)}{(\lambda+\alpha)^{2}}$ &$\frac{-1}{\alpha(\alpha+\lambda)}<\frac{2}{3}$\\
\hline
\end{tabular}
\end{center}
\caption[crit]{\label{tab2}
The properties of the critical points when $m=2$ in the phantom case.}
\end{table*}

\section{Conclusion}\label{sec4}
We consider radiation, matter and dark energy and the interactions between them, we then did a full
dynamical analysis for both quintessence and phantom cases. For the quintessence case,
we find that the unstable fixed points $S_{qR}$, $M_q$ and $S_{qRM}$ help weaken the coincidence problem.
In particular, $S_{qR}$ provides small $\Omega_\phi$ after the totally radiation dominated ($\Omega_r=1$) era,
$M_q$ provides the matter dominated era with nonzero $\Omega_\phi$, and $S_{qRM}$ helps
weaken the triple coincidence problem and prepares the universe
for the dark energy domination era $S_{qM}$. The interaction between dark matter and dark energy
helps alleviate the coincidence problem for the quintessence case. However, the interaction
cannot alleviate the coincidence problem for the phantom case. The interaction between radiation
and dark energy does not affect the dynamical behavior of the universe.


\acknowledgments
 This work was supported by  China Postdoc Grant
No.20100470237£¬ the Ministry of Science and Technology of China
national basic science Program (973 Project) under grant Nos.
2007CB815401 and 2010CB833004, the National Natural Science
Foundation of China key project under grant Nos. 11005164
and 10935013, and the Distinguished Young Scholar Grant 10825313,
CQ CSTC under grant
No. 2009BA4050 and 2010BB0408, and CQ MEC under grant No. KJTD201016.

\end{document}